\documentclass[twocolumn,showpacs,preprintnumbers,amsmath,amssymb]{revtex4}
\pdfoutput=1

\usepackage{graphicx}
\usepackage{dcolumn}
\usepackage{bm}
\usepackage{latexsym}

\begin{document}

\preprint{APS/123-QED}

\title{A Theory for Particle Settling and Shear-Induced Migration in Thin Film Flow}

\author{Benjamin P. Cook}
\affiliation{Mathematics Department, University of California Los Angeles \\ Los Angeles, California}
\email{bencook@math.ucla.edu}

\date{\today}

\begin{abstract}
Experiments of particle-laden inclined film flow [Zhou, Dupuy, Bertozzi, and Hosoi, Phys. Rev. Lett. 94 (2005)] have displayed different settling behaviors depending on the particle concentration $\phi$ and angle of inclination $\theta$, in which particles accumulate on the substrate or near the advancing contact line, or remain mixed.  Zhou et al. presented a lubrication model that captures the qualitative behavior of the high-$\phi$, high-$\theta$ regime, characterized by a particle-rich ridge near the contact line, but cannot explain the other observed settling behaviors.  This work presents a model in which $\phi$ varies through the film depth, unlike Zhou et al.'s model.  Average velocities for the liquid and particulate phases are computed, and the implications for phase separation are discussed.  It is found that the equilibrium depth profile of $\phi$ is more important than gravitational settling in the down-slope direction in determining phase separation.  The predicted settling behavior is directly compared with  Zhou et al.'s experimental data.

\end{abstract}

\pacs{47.15.gm, 47.55.Kf, 47.55.nd, 47.57.ef}
\maketitle

\section{Introduction}

\begin{figure}
\includegraphics[width=3.3in]{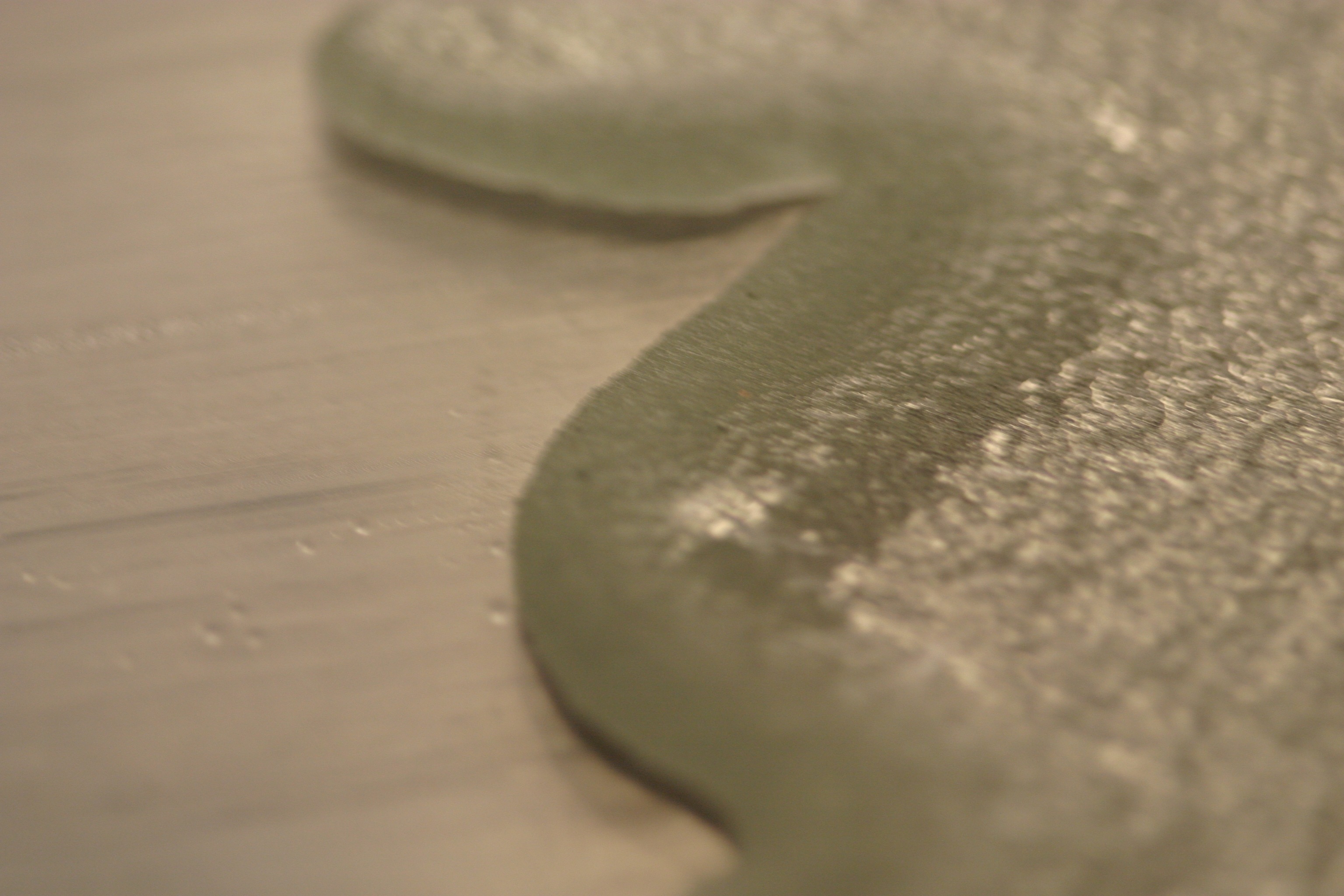}
\caption{\label{photo} Particle-rich ridge in an inclined film experiment.}
\end{figure}

Film flow of particle-laden liquid occurs in many important contexts, from geophysical flows such as erosion and turbidity currents \cite{hutter} to industrial processes including papermaking and the application of fertilizers.  While sophisticated constitutive models have been developed for general suspension flow, these models are generally not compatible with the lubrication approximation used for single-phase film flow.  As a result, the mathematical description of particle-laden films remains a challenging problem.

The complexity of such films is evident in a recent study by Zhou et al. \cite{ZDBH} of flow on an incline.  They observed three distinct flow types, characterized by the relative motion of the liquid and particulate phases.  At low inclination angles $\alpha$ and particle volume fractions $\phi$ the particles settle to the bottom substrate and are removed from the flow.  At intermediate $\alpha$ and $\phi$ the suspension appeared well mixed for the duration of their experiment.  At larger $\alpha$ and $\phi$ the particles were observed to accumulate near the advancing contact line, forming a pronounced ridge up to several times thicker than the upstream film.  They also reported that the growth of the fingering instability, which is known to deform the contact line in many film problems, is somewhat suppressed in the third regime.

Zhou et al. also introduced a lubrication model for this unique particle-rich ridge regime, which was revised and analyzed in \cite{cook}.  This model attributes the aggregation to the buoyant force on the denser particles, with the relative velocity specified by a hindered settling function $f(\phi)$ \cite{rz}.  The bulk motion is determined by balancing the gravity force with the viscous stress, which is expressed in terms of an effective viscosity $\mu(\phi)$ \cite{krieger}.  Thus particles move downstream slightly faster than the fluid and accumulate near the contact line, where the larger viscosity makes the film thicker.

In order to depth-average the Stokes equations for the lubrication approximation, the dependence of $\phi$ on the coordinate $z$ normal to the plane must be specified.  The model of Zhou et al. assumes $\phi$ is independent of $z$, which prohibits particles from settling to the substrate.  They use this assumption for simplicity, but note it is reasonable because shear-induced diffusion can oppose the settling of particles in the $z$ direction, possibly resulting in a stable depth profile.  Such an equilibrium is the likely explanation why particles do not always settle out of the flow as in the low-$\alpha$, low-$\phi$ regime.  Zhou's assumption has the consequence that the particle-rich ridge by the mechanism described above occurs in the lubrication model regardless of $\alpha$ and $\phi$.  

This letter proposes a model that balances shear-induced migration with settling to explicitly determine the depth profile of $\phi$, which is necessary in order to describe particles settling to the substrate and to explain the existence of distinct settling behaviors.  The specific depth profile also has a large impact on the relative velocities of the two phases: the phase-averaged velocity of the mixture depends strongly on $z$, so a given phase will move faster when it is concentrated near the free surface rather than near the substrate.  It is found that at an equilibrium profile this effect represents a larger contribution to the relative velocity than does settling in the flow direction, which suggests a three-dimensional treatment reflecting the stratified nature of the flow may be necessary to accurately describe the ridge phenomenon.  

Similar models have been studied before, notably by Schaflinger et al. \cite{schaflinger} and Timberlake and Morris \cite{morris}.  Shaflinger et al. used the ``diffusive flux" model for shear-induced diffusion introduced by Leighton and Acrivos \cite{migration}, which states that the volume flux of particles is given by 
\begin{equation} \label{leighton}
N_d = -a^2 \dot{\gamma} \hat{D}(\phi) \nabla \phi,
\end{equation}
where $\dot{\gamma}$ is the shear rate, $a$ is the particle radius, and the dimensionless diffusion coefficient was found by Leighton \cite{leighton} to be well approximated by $\hat{D}(\phi) = \frac{1}{3} \phi^2 (1+\frac{1}{2} e^{8.8 \phi})$.  The use of the scalar shear rate restricts this model to simple shear flows, which nonetheless include film flow where $\dot{\gamma}=dv/dz$ and $v$ is the velocity of the mixture.  Schaflinger et al. balanced this flux with that due to gravitational settling in the $z$ direction, which they approximated with a hindered settling function.  This condition along with the Newtonian stress balance allowed them to derive a system of two first-order ordinary differential equations for the concentration and shear stress, which they solved numerically.  

Two important features of the solutions can be deduced from the form of \eqref{leighton}.  Because the flux is proportional to the shear rate, the vanishing stress at the free surface $z=h$ ensures there is no diffusive flux to balance settling, and therefore $\phi(h)=0$ for all solutions\footnote{A steady solution also requires no diffusive flux where the maximum concentration $\phi_m$ is reached, corresponding to packed spheres, however this cannot happen at the free surface in Shaflinger et al.'s model because $d\phi/dz \leq 0$.}.  Also, the diffusive flux must be always directed upward in order to balance gravity, which implies by \eqref{leighton} that $d\phi/dz \leq 0$.

Timberlake and Morris included theory for the depth profile of concentration in their experimental paper on film flow of a neutrally buoyant suspension.  Their description uses the ``suspension balance" model of Nott and Brady \cite{nott-brady} for particle migration.  That more rigorous model calculates a ``temperature" measuring fluctuations in particle velocities, which is generated by shear, dissipated by viscous stress, and diffuses through an effect related to the finite particle size.  This last property is the most significant difference between the diffusive flux and suspension balance models, implying that particle migration depends nonlocally on the shear rate, which in this case allows a small nonzero concentration at the free surface.  Otherwise the two models generally give similar predictions \cite{fang}.  Since Timberlake and Morris considered neutrally buoyant particles, $\phi$ increases with $z$, which is also confirmed by their experiment.  Rather than assuming the film is always in diffusive equilibrium, they retain the $x$ coordinate in the flow direction, and their calculations indicate a distance on the order of $200h$ is necessary to reach equilibrium.  This factor decreases strongly with the bulk concentration and is proportional to $(h/a)^2$.

\section{model}

This work will use the diffusive flux model for simplicity, and proceed similarly to Schaflinger et al., but differ crucially by using an extra term in which the particle flux opposes gradients in the shear rate, in addition to opposing concentration gradients as in \eqref{leighton}.  This effect was introduced in \cite{migration} and quantified in \cite{phillips} in the expression
\begin{equation} \label{phillips}
\frac{D\phi}{Dt} = a^2 \nabla \cdot \left[ K_c \phi \nabla(\dot{\gamma} \phi) + K_\eta \dot{\gamma} \frac{\phi^2}{\mu(\phi)} \nabla \mu(\phi) \right]
\end{equation}
for the particle migration, where the best fit with experiment was obtained with the values $K_c=0.43$ and $K_\eta=0.65$ for the two constants.  Equation \eqref{phillips} corresponds to a particle flux
\begin{eqnarray} \label{phillips sigma}
F_m = -a^2 K_c \phi \nabla \left(\frac{\sigma}{\mu(\phi)} \phi \right) - a^2 (K_\eta - K_c) \frac{\sigma \phi^2}{\mu(\phi)^2} \nabla \mu(\phi) \nonumber \\
= -\frac{a^2 \phi}{\mu(\phi)} \left( K_c \nabla(\sigma \phi) + (K_\eta-K_c) \frac{\sigma \phi}{\mu(\phi)} \nabla \mu(\phi) \right),
\end{eqnarray}
where the shear rate $\dot{\gamma}$ has been eliminated in favor of the shear stress $\sigma = \mu(\phi) \dot{\gamma}$.

For a flat film on an incline, equilibrium is reached when this flux balances that of gravitational settling in the $z$ direction.  Settling rates are commonly expressed as a product of the velocity of a single sphere $v_s = -(2/9) \Delta \rho g/\mu_f$ by a hindered settling function $f(\phi)$ for which many empirical formulas exist.  Here $\rho$ and $\mu_f$ are the density and viscosity of the fluid, $g$ is the gravitational constant, and $\Delta=(\rho_p-\rho)/\rho$ is the density difference for particles of density $\rho_p$.  In this case it is convenient to follow Schaflinger et al. and use the hindered settling function $f(\phi)=(1-\phi)/\mu(\phi)$, leading to the settling flux
\begin{equation} \label{settling flux}
F_s = -\frac{2}{9} \frac{a^2 \Delta \rho g \cos \alpha}{\mu_f} \frac{\phi (1-\phi)}{\mu(\phi)},
\end{equation}
where $\alpha$ is the angle of inclination.

The balance of flux $F_m+F_s=0$ then takes the form
\begin{equation} \label{flux balance 1}
K_c (\sigma \phi)' + (K_\eta-K_c) \frac{\sigma \phi}{\mu(\phi)} \mu(\phi)' = -\frac{2}{9} \frac{ \Delta \rho g \cos \alpha}{\mu_f} (1-\phi)
\end{equation}
where the gradients have been replaced primes denoting differentiation by $z$.  Substituting the standard formula $\mu(\phi) = \mu_f (1-\phi/\phi_m)^{-2}$  \cite{krieger} with the maximum packing fraction $\phi_m \approx 0.67$ and differentiating yields
\begin{eqnarray} \label{flux balance 2}
\left[1 + \frac{2(K_\eta-K_c)}{K_c} \frac{\phi}{\phi_m-\phi} \right] \sigma \phi' \nonumber \\ 
= \phi(1+\Delta \phi) - \frac{2 \Delta}{9 K_c} (\cot \alpha) (1-\phi),
\end{eqnarray}
where $z$ and $\sigma$ have now been nondimensionalized using the depth of the film $h$ and the unit of stress $(\rho g/h) \sin \alpha$.

For a flat film there is no capillary force, so the pressure can be set to zero at the free surface $z=1$, and is assumed to be hydrostatic in the suspension.  The nondimensional shear stress then satisfies the equation
\begin{equation} \label{stress}
\sigma' = -(1+\Delta \phi).
\end{equation}
Equations \eqref{flux balance 2} and \eqref{stress} constitute the system to be studied here, with the understanding that \eqref{flux balance 2} is replaced by $\phi'=0$ when $\phi=0$ or $\phi=\phi_m$ to ensure pure fluid and packed particles are admissible solutions and to keep the concentration within its meaningful range.  The physical boundary conditions both involve the stress:  $\sigma(0)=(1+\Delta \phi_0)$ and $\sigma(1)=0$, where $\phi_0$ is the imposed average concentration.  Thus for these two equations there is only a one-parameter family of physically meaningful solutions, parameterized by $\phi_0$.  In practice this system was easiest to solve by shooting with a Runge-Kutta method from $z=0$ while adjusting the value of $\phi(0)$.  Once $\sigma$ and $\phi$ are determined, the mixture velocity can be calculated using $dv/dz=\dot{\gamma}=\sigma(z)/\mu(\phi(z))$ and $v(0)=0$. 

\section{solutions}

\begin{figure}
\includegraphics[width=3.3in]{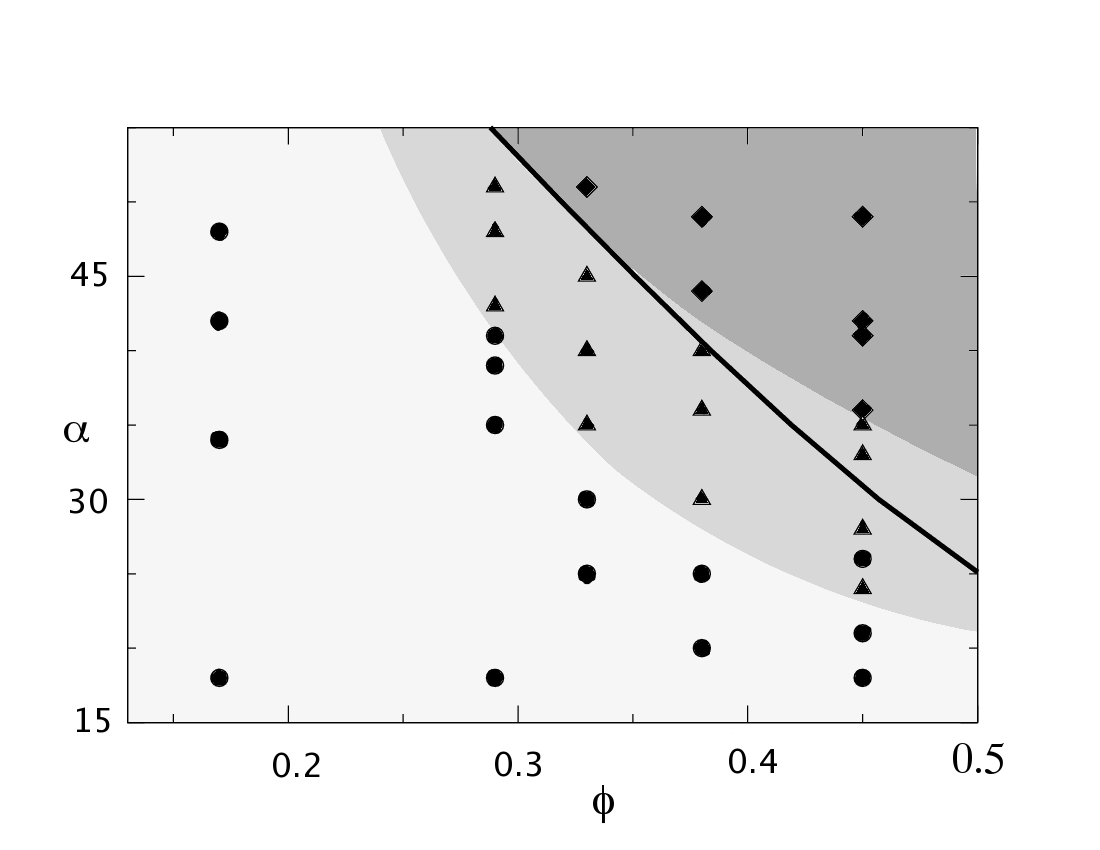}
\caption{\label{overlay} The function $\phi^*(\alpha)$ determining whether particles tend toward the top or bottom of the film.  Overlaid are Zhou et al.'s experimental parameters for which particles settle to the substrate ($\bigcirc$, white), remain well mixed ($\bigtriangleup$, light), or accumulate in a ridge ($\Diamond$, dark).  Experimental data are from figure 2 of \cite{ZDBH}.}
\end{figure}

Since particle migration in this model does not strictly oppose the concentration gradient, $\phi$ is not constrained to decrease with $z$ as in the work of Schaflinger et al.  The lack of a migration flux at the free surface however is general to the diffusive flux model, and still applies here, forcing either $\phi(1)=0$ or $\phi(1)=\phi_m$.  Since $\sigma \geq 0$, it is also apparent from equation \eqref{flux balance 2} that $\phi(z)$ is monotone, because $\sigma \phi'$ is determined by a function of $\phi$ only with a single unstable root $\phi^* = \phi^*(\alpha)$ in its allowable domain (between $0$ and $\phi_m$).  There are then two possibilities: $\phi_0 > \phi(0) > \phi^*$ with $\phi(1) = \phi_m$, or $\phi_0 < \phi(0) < \phi_m$ with $\phi(1) = 0$.

In the latter case, the particulate phase is located preferentially near the bottom of the film and (because $v(z)$ is always increasing) moves slower than the fluid on average, both of which are necessary conditions for the particles to settle out of the flow.  It seems natural then to associate $\phi_0<\phi^*(\alpha)$ with this regime in Zhou et al.'s experimental work \cite{ZDBH}.  The case $\phi_0>\phi^*(\alpha)$ should then correspond to the particle-rich ridge regime, as the particles do not settle to the bottom and move faster on average than the fluid, even without including the settling velocity in the flow direction.  While there is no obvious reason why there should be a regime (other than the single solution $\phi \equiv \phi^*$) where the fluid and particles move at the same velocity, it may be that experiments in which the suspension stayed well-mixed had $\phi_0 \approx \phi^*$ and the relatively small difference between the two velocities did not have time to produce noticeable segregation on the experimental time scale.  

\begin{figure}
\includegraphics[width=3.3in]{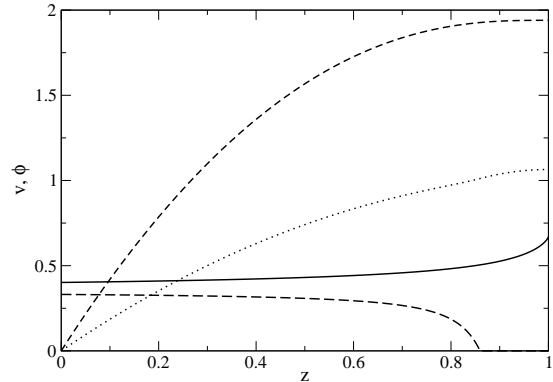}
\caption{\label{sample} Depth profiles of $\phi$ and $v$ for two average concentrations at $\alpha=45^\circ$.  Bulk concentration $\phi_0=0.25$:  velocity (dot) and concentration (long dash), bulk concentration $\phi_0=0.45$:  velocity (short dash) and concentration (solid).  Velocities are scaled by the average velocity of a homogeneous film at the same concentration.  With this rescaling the average velocities at $\phi_0=0.25$  of the particle and liquid phases are $0.57$ and $0.70$, and at $\phi_0=0.45$ the velocities are $1.41$ and $1.33$ respectively.}
\end{figure}

Plotted in figure \ref{overlay} is the calculated transition point $\phi^*(\alpha)$ and the experimental data from \cite{ZDBH}.  As expected, the transition lies within the well-mixed regime.  This calculation involves no fitting parameters, and the agreement is remarkable considering the simplifying assumptions of one-dimensional, time-independent flow.  The position of the curve $\phi^*(\alpha)$ also suggests that the experimentally observed well-mixed films mostly lie in the $\phi_0 < \phi^*(\alpha)$ range, and therefore would likely result in particles settling out of the flow were the experiments continued longer.

Examples of the two cases ($\phi_0 > \phi^*$ and $\phi_0 < \phi^*$) are shown in figure \ref{sample} for $\alpha=45^\circ$, $\phi^*(\alpha) \approx 0.35$.  The effect of the increasing concentration profile for $\phi_0=0.45$ is to flatten the velocity near the top from the parabolic shape of an unstratified film, while for $\phi_0=0.25$ the absence of particles near the top increases the shear in this area.  Also of interest is the fact that when $d\phi/dz>0$ both phases move faster than the velocity of an unstratified film, because of the high-shear, low-$\phi$ region at the bottom and the low shear at the top where $v$ is at its greatest.  Both phases are slower when $d\phi/dz<0$.

\begin{figure}
\includegraphics[width=3.3in]{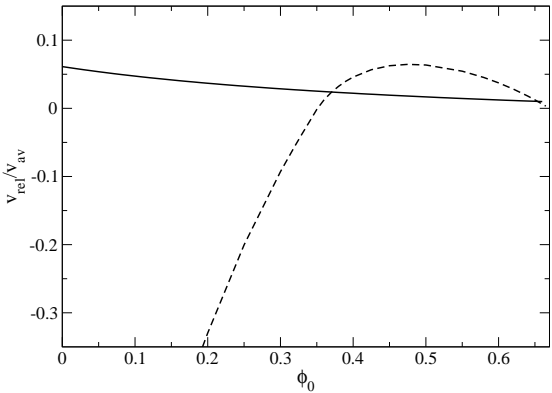}
\caption{\label{vrel} The ratio $v_{rel}/v_{av} = (v_p - v_f)/(\phi v_p + (1-\phi) v_f)$ of velocities relevant for formation of the particle rich ridge.  Velocity difference due to the stratified flow as described above (dash), and velocity difference due to direct gravitational settling in the flow direction as described by Zhou et al. (solid).}
\end{figure}

In figure \ref{vrel} the relative velocity due to stratification is compared with the in-plane settling velocity used in \cite{ZDBH} and \cite{cook} at $\alpha=45^\circ$.  Specifically the vertical axis measures the ratio $v_{rel}/v_{av} = (v_p - v_f)/(\phi v_p + (1-\phi) v_f)$ that determines the accumulation of particles in an experiment limited by the length of the channel.  For concentrations greater than $0.37$, stratification has a larger effect than in-plane settling.  Since the particle-rich ridge occurs at rather high concentrations, the stratified flow appears to be the more important cause of the ridge.

A description of the ridge evolution including stratification is possible within the lubrication context if the film is assumed to be always in equilibrium between settling and migration, by using the calculations of figure \ref{vrel} to determine the relative velocity from $\phi$.  This would result in a system similar to that in \cite{cook}, which for length scales greater than a modified capillary length describes a ridge that grows linearly with time.
If this route is followed, care must be taken to ensure the length scale is also large enough to justify the equilibrium assumption.
The experiments and two-dimensional calculations of Timberlake and Morris \cite{morris} indicate the distance travelled before reaching equilibrium can be as large as tens of centimeters, even for an experiment with fairly large particles such as \cite{ZDBH}.  
At shorter length scales, such a two-dimensional model may therefore be necessary, which would generalize the above results by allowing non-equilibrium concentration profiles.
The most likely effect of non-equilibrium physics would be to lengthen the timescale of phase separation, making the well-mixed regime more likely for length-limited experiments.

This new theory demonstrates the importance of particle migration in determining the flow, and can provide a starting point for studying effects such as ridge formation, particle deposition in the clear fluid regime, the contact-line instability, or span-wise particle banding \cite{carpen}.

This work is part of my doctoral dissertation at UCLA, and I am grateful to my advisor Andrea Bertozzi for her help and guidance.  Financial support was provided by NSF grants ACI-0321917 and DMS-0502315 and ONR grant N000140710431.


\bibliography{zbalance}

\end{document}